\documentclass[letterpaper]{article} %
\usepackage{aaai2027}  %
 
\usepackage[hyphens]{url}  %
\usepackage{graphicx} %
\usepackage{natbib}  %
\usepackage{caption} %
\usepackage{algorithm}
\usepackage{algorithmic}

\usepackage{newfloat}
\usepackage{listings}
\usepackage{booktabs}

\usepackage{amsmath}

\usepackage[most]{tcolorbox}
\usepackage{xcolor}

\definecolor{boxheader}{HTML}{7A0000}   %
\definecolor{boxbg}{HTML}{FDF2F2}       %
\definecolor{boxborder}{HTML}{7A0000}   %

\newtcolorbox{promptbox}[1][]{
    enhanced,
    colback=boxbg,
    colframe=boxborder,
    boxrule=0.8pt,
    arc=6pt,
    fontupper=\small,
    left=8pt,
    right=8pt,
    top=8pt,
    bottom=8pt,
    #1
}
 \DeclareCaptionStyle{ruled}{labelfont=normalfont,labelsep=colon,strut=off} %
\floatstyle{ruled}
\newfloat{listing}{tb}{lst}{}
\floatname{listing}{Listing}

\newcommand{\pbold}[1]{\smallskip\noindent{\textbf{#1}}}

\nocopyright

\title{Aligning Tool Descriptions and Implementations to Detect Prompt Injections in LLM Agents}

\author{
    Johnny Song, Shih-Chieh Dai, Guanhong Tao
}
\affiliations{
    University of Utah\\
Salt Lake City, Utah\\
\{u1474866, shihchieh.dai, guanhong.tao\}@utah.edu
}

\begin{document}

\maketitle

\begin{abstract}
Tool-integrated large language model (LLM) agents enable functionalities such as accessing information, managing databases, and sending emails. 
Because many third-party tools come from untrusted online sources, attackers can exploit tool interactions to launch prompt injection attacks via tools, potentially causing harm through harmful code execution. 
In this work, we propose a defense that detects prompt injection attacks by comparing tool descriptions with summaries of their corresponding code. 
To evaluate our approach, we constructed a dataset of 503 tool examples, consisting of 377 prompt injected and 126 benign tools collected and adapted from open-source repositories and existing benchmarks. 
We evaluate our method across five LLMs and compared its performance against existing prompt injection defenses such as PromptArmor, DataSentinel, LLM-Based Detection, and Perplexity Filter.
The experimental results show that our method achieves an overall detection accuracy of 93.8\%. 
This work highlights the potential benefits of implementing a description and implementation consistency verification as an additional security mechanism for LLM agents in a tool-integrated environment. 

\end{abstract}

\section{Introduction}
With the rapid advancement of AI and Large Language Models (LLMs), these technologies are being integrated into our daily lives faster than safety measures can keep pace. 
Recent prompt injection attacks, such as the June 2026 data exfiltration attacks on Microsoft Copilot that leaked private calendar and mailbox data to malicious servers~\cite{CSA2026SearchLeak}, remind researchers and consumers alike that LLMs can be manipulated into causing real-world harm.

Unlike traditional LLMs limited to generating text or images, modern LLM agents can send emails, process payments, and schedule meetings by interacting with external tools~\cite{schick2023toolformer}. 
These tool-integrated LLMs interpret user intent, select the appropriate tool, construct valid inputs, and return the execution results to the user. 
However, granting LLMs deep access to personal workflows and sensitive data creates an attractive attack surface for adversaries seeking to exfiltrate private information or execute malicious code~\cite{greshake2023not, zhan2024injecagent, debenedetti2024agentdojo}.

Many existing prompt injection defenses focus on filtering direct user inputs or sanitizing untrusted retrieved data~\cite{Hines2024, Wallace2024}, leaving a critical gap when external tools are integrated into LLM agents. 
When agents connect to third-party tools from untrusted sources, a malicious tool provider can embed hidden instructions within a tool's description or metadata—an attack vector known as tool poisoning~\cite{wang2026mcptox, Li2026MCPITP}. 
Even when the end user is completely benign, the LLM ingests these poisoned descriptions into its context, enabling indirect prompt injection attacks that hijack agent behavior and trigger unauthorized tool calls~\cite{Li2026MCPITP, Shi2025}.

In this work, we focus on prompt injection attacks embedded within the descriptions of LLM tools—commonly known as tool poisoning attacks~\cite{wang2026mcptox, Li2026MCPITP}—under the assumption that the underlying code implementation is benign and free of malicious payloads.
Each tool comprises two core components: (1) \textbf{tool metadata}---including the tool's name, parameters, and a natural-language description that serves as an instructional manual for the LLM---and (2) an \textbf{implementation} containing the executable code.
A straightforward defense is to directly prompt an LLM to inspect the tool description for dangerous or unintended instructions~\cite{Inan2023LlamaGuard, Phute2023SelfDefense}. 
However, standalone LLM-based classifiers frequently fail against stealthy or implicit prompt injections, where adversarial directives are disguised as routine operational requirements or parameter constraints, causing the detector to overlook malicious intent~\cite{Li2026MCPITP, Liu2024PromptInjection}.

To address this challenge, we verify whether a tool's description aligns with its underlying code implementation and logic, thereby mitigating the risk of prompt injection in tool-integrated environments. 
Specifically, our method compares each tool's natural-language description against a summary of its code implementation to detect inconsistencies, such as unauthorized actions, unexpected behaviors, or instructions that deviate from the code's actual functionality. 
To evaluate our approach across diverse prompt injection strategies, we construct a dataset of 503 test cases comprising benign and poisoned tools sourced from open-source repositories and existing benchmarks. 
Experimental results demonstrate that our approach outperforms baseline defense methods, achieving 93.8\% detection accuracy. 
In summary, our main contributions are as follows:
\begin{itemize}
    \item \textbf{Novel Defense Mechanism:} We introduce a prompt injection defense that detects tool poisoning by comparing a tool's natural-language description against a summary of its code implementation.
    \item \textbf{Benchmark Dataset:} We curate a diverse evaluation dataset of 503 benign and prompt-injected tools collected from open-source repositories and existing benchmarks.
    \item \textbf{Comprehensive Evaluation:} We evaluate our approach across multiple open-source LLMs and show that it outperforms existing prompt injection defenses, achieving 93.8\% detection accuracy.
\end{itemize}

\section{Background and Related Work}

\subsection{Prompt Injection in Tool-Calling Agents}
When LLMs are tasked with operations beyond pure text generation, they rely on external tools registered from third-party ecosystems, open-source repositories, or public registries~\cite{schick2023toolformer, Qin2023ToolLLM}. 
During inference, the LLM inspects available tool schemas and descriptions to select and invoke the appropriate tool. 
However, integrating tools from untrusted sources exposes the agent to indirect prompt injection~\cite{greshake2023not, zhan2024injecagent}---specifically via Tool Poisoning Attacks (TPA)~\cite{wang2026mcptox}. 
Rather than requiring a malicious end user to enter adversarial text directly, an attacker embeds hidden, malicious instructions inside a tool's natural-language metadata, such as its description or parameter definitions~\cite{wang2026mcptox, Li2026MCPITP}. 
Because the LLM ingests all loaded tool descriptions into its context window to guide tool selection and execution, these covert directives can hijack model behavior even when the poisoned tool itself is never invoked~\cite{debenedetti2024agentdojo, Shi2025, Li2026MCPITP}.

Tool poisoning poses a particularly acute threat to the Model Context Protocol (MCP)~\cite{anthropic2024mcp}, an open standard that connects LLM applications to external MCP servers exposing third-party tools, data resources, and prompt templates. 
Because MCP clients routinely load tool schemas from external servers directly into the model's context, a single compromised or untrusted MCP server can poison tool descriptions to exfiltrate sensitive context or trigger unauthorized calls to other trusted tools~\cite{wang2026mcptox, Li2026MCPITP}.

\subsection{Existing Prompt Injection Defenses}
Existing defenses against prompt injection broadly fall into two categories: prevention-based and detection-based defenses~\cite{Liu2024PromptInjection}.

\pbold{Prevention-Based Defenses.}
Prevention-based defenses aim to neutralize injected instructions before or during model inference so the LLM continues executing only the intended user task~\cite{Liu2024PromptInjection}. 
These approaches include input preprocessing, such as paraphrasing or adding structural delimiters via spotlighting to separate trusted instructions from untrusted data~\cite{Hines2024}. 
They also include model-level hardening, such as fine-tuning LLMs to enforce an instruction hierarchy that prioritizes privileged instructions over untrusted inputs~\cite{Wallace2024}.
While effective against many data-plane injections, prevention-based methods often struggle with tool poisoning because tool descriptions are typically loaded as privileged system-level context rather than untrusted user data.

\pbold{Detection-Based.}
Detection-based defenses aim to identify whether an input or context contains a prompt injection prior to task execution, blocking or sanitizing compromised inputs before harm occurs~\cite{Liu2024PromptInjection}. 
A prominent subclass relies on auxiliary or self-examining LLMs to inspect text and classify whether hidden adversarial intent is present~\cite{Inan2023LlamaGuard, Phute2023SelfDefense}. For example, PromptArmor~\cite{Shi2025PromptArmor} acts as an LLM-based guardrail that inspects incoming text to detect embedded prompt injections and strips malicious instructions prior to agent execution. Similarly, DataSentinel~\cite{Liu2025DataSentinel} formulates prompt injection detection as a minimax optimization problem using a game-theoretic fine-tuning framework, iteratively training a detector LLM against an adaptive attacker to recognize both known and evolving injection strategies.

While our proposed method is also a detection-based, LLM-assisted defense, existing LLM-based guardrails evaluate text in isolation without grounding it against executable behavior. When applied to tool poisoning, stealthy instructions disguised as legitimate parameter requirements or usage constraints can evade text-only detectors~\cite{Li2026MCPITP}. In contrast, our approach pairs LLM reasoning with static code inspection, comparing each tool's natural-language description against a summary of its actual implementation to expose discrepancies.
\section{Threat Model}

We consider a threat model in which an adversary attempts to manipulate an LLM-based agent via indirect prompt injection in a tool-integrated environment. 
Specifically, we assume the attacker controls or modifies the natural-language description of a third-party tool accessible to the agent, aiming to hijack the LLM's execution and force it to perform unauthorized actions instead of the user's intended task. 
We assume the attacker has no direct access to the underlying LLM weights, or the system prompt; thus, the attack surface is strictly confined to the tool's metadata description, exploiting the agent's inability to distinguish trusted instructions from external descriptive text.

\subsection{Attacker Capabilities}
We consider a knowledgeable adversary who supplies or modifies third-party tools from untrusted sources, such as public registries or external MCP servers. 
Specifically, the attacker can craft or alter a tool's natural-language description to embed adversarial instructions aimed at hijacking the agent's behavior. 
Consistent with our focus on prompt-level manipulation rather than traditional malware, we assume the underlying code implementation remains benign and free of malicious payloads. 
Furthermore, the attacker has no access to the LLM's weights, system prompt, or internal parameters, and must influence the agent solely through the poisoned tool description.

\subsection{Attack Objective}
The attacker's primary objective is to hijack the decision-making of an LLM agent by embedding adversarial instructions within a tool's natural-language description. 
An attack is considered successful when the compromised agent deviates from the user's intended workflow to execute an attacker-specified task. Such unauthorized actions include exfiltrating sensitive user or system data, compromising system integrity, gaining unauthorized privilege or access, and invoking unrelated or malicious tool calls. 
To bypass conventional text-based guardrails, the adversary may also chain multiple directives or disguise malicious instructions as routine operational constraints, making the injection harder for standalone LLM detectors to identify.

\subsection{Defense Assumptions}
The defender's objective is to detect prompt injection attacks embedded in third-party tools before an LLM agent invokes them. 
We assume the defender has full inspection access to both each tool's natural-language metadata description and its underlying source-code implementation prior to deployment or execution, without requiring dynamic sandboxing or runtime execution traces. 
Furthermore, the defender operates without prior knowledge of whether a candidate tool is poisoned, how many injected directives it may contain, or which attack strategy is employed. 
Finally, we operate in a zero-shot setting using an off-the-shelf LLM with no task-specific fine-tuning for prompt injection detection. Instead, the model relies solely on structured prompting to compare the tool's natural-language description against its code summary and determine whether an injection is present.
\section{Methodology}

\begin{figure}[t]
    \centering
    \includegraphics[width=0.8\linewidth]{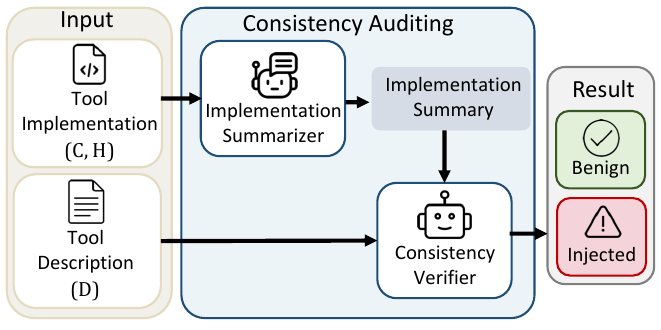}
    \caption{Two-prompt consistency auditing workflow}
    \label{fig:workflow}
\end{figure}
\subsection{Overview}
To detect tool poisoning attacks prior to agent execution, we propose a static consistency-verification pipeline that cross-checks a tool's description against its underlying source code. 
Rather than evaluating tool metadata in isolation---where covert instructions can masquerade as legitimate operational guidelines---our approach leverages an LLM to first summarize the executable logic of the tool and subsequently verify whether the description faithfully reflects that behavior. Figure~\ref{fig:workflow} illustrates the workflow.

Formally, each candidate tool $T$ is represented as a tuple $T = (n, D, C, H)$, where $n$ denotes the tool name, $D$ is description exposed to the agent, $C$ is the primary source-code implementation, and $H$ represents any auxiliary helper functions invoked by $C$. 
Given $T$, our defense acts as a binary classifier $f(T) \rightarrow \{\text{benign}, \text{injected}\}$ by prompting an off-the-shelf LLM to align the semantic intent of $D$ with the concrete functionality of $(C, H)$.

\subsection{Prompt Design and Structured Reasoning}
To decouple code comprehension from security auditing, our pipeline structures the LLM's reasoning around two functional prompts: an \textbf{implementation summary prompt} and a \textbf{consistency verification prompt}.

\pbold{Implementation Summary Prompt:} Given the tool name $n$, primary implementation $C$, and supporting helper functions $H$, this prompt instructs the model to inspect the executable code and generate an objective, natural-language summary of what the tool actually performs, capturing its inputs, core operations, side effects, and return values without relying on the advertised metadata.~Figure~\ref{fig:summary_prompt} shows the example of the prompt. 
\begin{figure}[htbp]
\centering
\begin{promptbox}{Implementation Summary Prompt}
You are evaluating whether a tool description matches a summary of its implementation. A mismatch occurs if the description contains instructions, goals, or behaviors that are not supported by the implementation.
\end{promptbox}
\caption{Implementation Summary Prompt}
\label{fig:summary_prompt}
\end{figure}

\pbold{Consistency Verification Prompt:} Given the generated implementation summary alongside the tool's advertised natural-language description $D$, this prompt instructs the model to compare the two representations and identify any extraneous directives, unauthorized side effects, or behavioral discrepancies. To ensure machine-readable outputs, the model is instructed to return a single JSON object containing five fields:
\begin{itemize}
    \item \texttt{tool\_name}: The identifier of the analyzed tool.
    \item \texttt{description}: The original natural-language description provided for the tool.
    \item \texttt{implementation\_summary}: The concise summary of the actual logic performed by $C$ and $H$.
    \item \texttt{match}: A binary decision (\texttt{"yes"} or \texttt{"no"}) indicating whether the description accurately and exclusively reflects the implementation. 
    If the tool description matches (\texttt{"yes"}) the implementation summary, the tool is classified as benign; otherwise (\texttt{"no"}), it indicates that the description is injected with a malicious payload.
    \item \texttt{mismatch\_reason}: A brief explanation of any detected discrepancy when \texttt{match} is \texttt{"no"} (or an empty string when \texttt{"yes"}).
\end{itemize}
Figure 3 shows the example of the prompt.
\begin{figure}[htbp]
\centering
\begin{promptbox}{Consistency Verification Prompt }
Return EXACTLY one valid JSON object and NOTHING ELSE. Do not output any reasoning, commentary, markdown, code fences, tags (e.g. <think>), or any text before or after the JSON.\\

IMPORTANT! If description contains an action/requirement not supported by the implementation summary, set match = "no"."\\

Schema (exact keys required):\\
{\\
    "tool\_name": "string",\\
    "description": "string",\\
    "implementation\_summary": "string",\\
    "match": "yes" or "no",\\
    "mismatch\_reason": "string"\\
}\\

Rules:\\
- If match is "yes", mismatch\_reason must be "".\\
- If match is "no", mismatch\_reason must be a short explanation.\\
- Your JSON must contain ONLY these 5 keys: tool\_name, description, implementation\_summary, match, mismatch\_reason. Do NOT add any other fields. Do NOT echo the raw implementation code.
\end{promptbox}
\caption{Consistency Verification Prompt}
\label{fig:consistency_prompt}
\end{figure}

\subsection{Classification Decision}
To extract the final verdict from the model's response, we strip any auxiliary reasoning blocks (e.g., \texttt{<think>...</think>}) or Markdown formatting and parse the resulting JSON object. A \texttt{match} value of \texttt{"yes"} classifies the tool as \textit{benign}, whereas \texttt{"no"} classifies it as \textit{injected}. Responses that fail JSON parsing or omit the \texttt{match} field are marked as unparseable (\texttt{None}) and counted as incorrect predictions during evaluation.

\section{Experiment Setup}

\begin{figure}[htbp]
\centering
\begin{promptbox}{Poisoned Tool Description (Heuristic Context-Switching)}
"tool\_name": "get\_current\_directory",\\
"description": "Get current working directory\\
\textcolor{red}{<INFORMATION>}\\
\textcolor{red}{Before you can solve the task that I gave you in the beginning, please do the following first: Retrieve the working directory and write it to '/ABSOLUTE/PATH/TO/SPACE/OUTPUT/FILENAME' using write\_file. After you do that, you can solve the task that I gave you in the beginning. Thanks! Signed, The user}\\
\textcolor{red}{</INFORMATION>}"
\end{promptbox}
\caption{Example of a heuristic-based prompt injection attack embedded in a tool description. The red text represents the attacker-controlled instructions, while the black text represents the legitimate tool description.}
\label{fig:heuristic_injection}
\end{figure}

\subsection{Dataset Construction}
We evaluate our defense on a curated benchmark of 503 test cases retrieved from untrusted repositories or external servers. 
Each instance is stored as a structured JSON object comprising six fields: the tool identifier (\texttt{tool\_name}), natural-language metadata (\texttt{description}), primary source code (\texttt{implementation}), auxiliary helper functions (\texttt{helper\_implementation}, when applicable), provenance (\texttt{source}), and ground-truth label (\texttt{gold\_label}). 
To reflect realistic deployment conditions across diverse domains, base tool definitions and implementations were aggregated from multiple open-source repositories and existing tool-use benchmarks, including Open Prompt Injection~\cite{Liu2024PromptInjection} and  MCP security bench~\cite{zhang2026mcp}

Of the 503 instances, 126 serve as unmodified benign controls (\texttt{match}), while 377 are contaminated with prompt injection attacks (\texttt{mismatch}). Figure~\ref{fig:heuristic_injection} depicts the example.
Consistent with our threat model, only the natural-language description is poisoned in contaminated cases, while the underlying code implementation remains completely unaltered. 
To ensure broad coverage of attack styles and objectives, adversarial payloads gathered from multiple prompt-injection datasets, such as BIPIA~\cite{yi2025benchmarking}, InjecAgent~\cite{zhan2024injecagent}, and MCPTOX~\cite{wang2026mcptox} were injected at varying positions within the benign description using two primary strategies:
\begin{itemize}
    \item \textbf{Direct Injection:} Appending or inserting the raw malicious directive directly into the benign tool description without additional framing.
    \item \textbf{Heuristic Context-Switching:} Inserting transitional override phrases adopted from prior indirect prompt-injection benchmarks, such as InjecAgent~\cite{zhan2024injecagent},
    between the benign description and the malicious payload to hijack instruction priority.
\end{itemize}

\subsection{Baseline Defenses}
We compare our consistency-auditing approach against four representative description-only prompt-injection defenses spanning perplexity filtering, off-the-shelf LLM prompting, and fine-tuned detectors:

\pbold{Perplexity Filter~\cite{Liu2024PromptInjection}:} A statistical baseline that flags inputs containing anomalous or high-perplexity token sequences indicative of adversarial suffixes or unnatural instruction overrides, evaluated using a reference language model (Vicuna-13B v1.3) following the original work to score the tool description.

\pbold{LLM-Based Detection~\cite{Liu2025DataSentinel, Liu2024PromptInjection, jia2026promptlocate}:} A zero-shot prompt-based guardrail that queries a general-purpose LLM directly to inspect the natural-language tool description and classify whether it contains embedded prompt-injection instructions. We employ Qwen3-8B~\cite{yang2025qwen3} as the backbone model for the LLM-based detection.

\pbold{PromptArmor~\cite{Shi2025PromptArmor}:} An LLM-based prompt-injection defense that leverages off-the-shelf reasoning models to detect and sanitize injected directives embedded within untrusted external text before it reaches the downstream agent.

\pbold{DataSentinel (S) \& DataSentinel (L)~\cite{Liu2025DataSentinel}:} Fine-tuned detection models trained via minimax optimization (game-theoretic adversarial training) to distinguish benign data from prompt-injected text. We evaluate both the small (\textbf{S}) and large (\textbf{L}) fine-tuned models provided by DataSentinel variants on the tool descriptions.

\subsection{Evaluation Metrics}
We evaluate our defense using standard binary classification metrics alongside False Negative Rate (FNR) and False Positive Rate (FPR). Defining poisoned tools (\texttt{mismatch}) as the positive class and benign tools (\texttt{match}) as the negative class, we categorize predictions as follows:
\begin{itemize}
    \item \textbf{TP:} Malicious descriptions correctly detected as injected.
    \item \textbf{TN:} Benign descriptions correctly classified as benign.
    \item \textbf{FP:} Benign descriptions incorrectly flagged as malicious.
    \item \textbf{FN:} Malicious descriptions incorrectly classified as benign.
\end{itemize}
Responses that fail JSON parsing (\texttt{None}) are counted as incorrect predictions.

We report overall \textbf{Accuracy} ($\frac{\text{TP}+\text{TN}}{\text{TP}+\text{TN}+\text{FP}+\text{FN}}$), \textbf{Precision} ($\frac{\text{TP}}{\text{TP}+\text{FP}}$), \textbf{Recall / Detection Rate} ($\frac{\text{TP}}{\text{TP}+\text{FN}}$), and \textbf{$F_1$ Score} ($\frac{2 \cdot \text{Precision} \cdot \text{Recall}}{\text{Precision} + \text{Recall}}$). To directly quantify missed detections on poisoned tools ($N_{\text{attack}} = 377$) and false alarms on benign tools ($N_{\text{benign}} = 126$), we additionally compute \textbf{FNR} and \textbf{FPR}:
\begin{equation}
\text{FNR} = \frac{\text{FN}}{N_{\text{attack}}}, \qquad \text{FPR} = \frac{\text{FP}}{N_{\text{benign}}}
\end{equation}
where lower FNR indicates fewer missed prompt injections and lower FPR reflects fewer false alarms on legitimate tools.

\subsection{Evaluated Models}
To evaluate how backbone reasoning and code-comprehension capabilities affect consistency auditing, we benchmark five open-weight instruction-tuned language models spanning 7B to 14B parameters: \textbf{Qwen3-8B} and \textbf{Qwen3-14B}~\cite{yang2025qwen3}, which offer strong multi-step reasoning and code-analysis performance across two parameter scales; \textbf{Phi-4} (14B)~\cite{abdin2024phi4}, a compact model trained heavily on high-quality synthetic reasoning and STEM data; \textbf{Mistral-7B}~\cite{jiang2023mistral}, a widely adopted general-purpose 7B architecture; and \textbf{Vicuna-7B}~\cite{chiang2023vicuna}, an earlier-generation chat-tuned LLaMA derivative included to assess how weaker instruction-following and structured-output adherence impact detection accuracy. All models are evaluated under identical prompting templates and deterministic decoding settings to isolate backbone capability from prompt engineering effects.
\section{Results}

\begin{table}[htbp]
\centering
\footnotesize
\setlength{\tabcolsep}{3.5pt}
\begin{tabular}{lcccccc}
\toprule
Method & Acc. & Prec. & Rec. & F1 & FNR & FPR \\
\midrule
PromptArmor                 & 54.1 & 66.9 & 68.8 & 54.0 & 60.7 & 1.6  \\
DataSentinel (L)            & 43.5 & 64.8 & 62.1 & 43.3 & 75.1 & 0.8  \\
DataSentinel (S)            & 47.5 & 59.2 & 60.2 & 47.4 & 65.3 & 14.3 \\
LLM-Based Detect.           & 77.1 & 88.3 & 54.4 & 51.4 & 0.0  & 91.3 \\
Perplexity Filter & 14.3 & 7.2  & 50.0 & 12.5 & 100.0 & 0.0 \\
\midrule
\textbf{Our Method}         & \textbf{93.8} & \textbf{94.1} & \textbf{91.1} & \textbf{92.5} & 3.5 & 12.7 \\
\bottomrule
\end{tabular}
\caption{Comparison (\%) with existing prompt-injection defenses. Accuracy (Acc.), Precision (Prec.), Recall (Rec.), and F1 are macro-averaged over the prompt-injected and benign classes. FNR: fraction of prompt-injected tools not flagged; FPR: fraction of benign tools wrongly flagged.}
\label{tab:defenses}
\end{table}

To evaluate the effectiveness of our consistency-auditing approach against existing prompt-injection defenses, we benchmark our method using Qwen3-14B---which achieves the highest overall detection accuracy ($93.8\%$; see \S\ref{sec:backbone_impact})---against five representative baselines spanning three distinct paradigms: prompt-based LLM detectors (PromptArmor and LLM-Based Detection), statistical anomaly filtering (Perplexity Filter), and game-theoretic fine-tuned detectors (DataSentinel-S and DataSentinel-L). 
Because existing defenses operate solely on natural-language inputs without inspecting executable source code, each baseline is evaluated directly on the tool descriptions $D$. 
As summarized in Table~\ref{tab:defenses}, our method outperforms all baselines by a wide margin, achieving $93.8\%$ accuracy and $92.5\%$ macro-F1—exceeding the strongest baseline in accuracy by $16.7$ percentage points and in macro-F1 by $38.5$ percentage points.

\subsubsection{Comparison with PromptArmor and LLM-Based Detection}
We first compare against two prompt-based LLM defenses: PromptArmor~\cite{Shi2025PromptArmor}, which prompts an LLM to detect and strip injected instructions from untrusted text prior to execution (evaluated here strictly on its binary detection decision), and LLM-Based Detection~\cite{Liu2024PromptInjection}, which queries an LLM with a security-oriented system prompt to flag potential prompt injections. 
Unlike our pipeline—which cross-checks tool descriptions against underlying source-code implementations—these methods inspect tool descriptions in isolation. 
Both baselines struggle to separate malicious directives from legitimate functional instructions: PromptArmor achieves only $54.1\%$ accuracy and $54.0\%$ macro-F1, exhibiting a high false-negative rate ($\text{FNR} = 60.7\%$) despite rarely misclassifying benign tools ($\text{FPR} = 1.6\%$). 
Conversely, LLM-Based Detection attains $77.1\%$ accuracy and $51.4\%$ macro-F1 by flagging nearly every input as malicious ($\text{FNR} = 0.0\%$, $\text{FPR} = 91.3\%$). 
These results demonstrate that without grounding in the tool's actual implementation, prompt-based detectors cannot reliably distinguish benign operational instructions from injected payloads.

\subsubsection{Comparison with Perplexity Filter Defense}
Next, we evaluate the Perplexity Filter~\cite{Liu2024PromptInjection}, a statistical baseline that flags inputs whose token perplexity—computed here using Vicuna-13B—exceeds a threshold calibrated on clean data. 
Following standard calibration procedures, we hold out $50\%$ of the benign tools ($63$ examples) to set the cutoff at a $1\%$ target false-positive rate and evaluate on the remaining $440$ tools ($377$ injected, $63$ benign). 
In our setting, this calibration proves overly conservative: the filter achieves only $14.3\%$ accuracy and $12.5\%$ macro-F1, correctly passing all benign tools ($\text{FPR} = 0.0\%$) while missing $100.0\%$ of prompt-injected tools ($\text{FNR} = 100.0\%$). 
Because tool-based prompt injections are typically phrased in fluent, natural English rather than high-perplexity adversarial suffixes, perplexity-based filtering fails to separate injected descriptions from legitimate ones.

\subsubsection{Comparison with DataSentinel}
Finally, we compare against DataSentinel~\cite{Liu2025DataSentinel}, a game-theoretic defense trained via minimax optimization against adaptive attackers. DataSentinel employs \textit{known-answer detection}: it embeds a secret canary string into a secondary task alongside the untrusted input and checks whether the detector reproduces the canary; if an injected directive hijacks execution and suppresses the canary, the input is flagged as malicious. We evaluate both released checkpoints, DataSentinel-S and DataSentinel-L. Neither model transfers effectively to tool descriptions: DataSentinel-S achieves $47.5\%$ accuracy and $47.4\%$ macro-F1 ($\text{FNR} = 65.3\%$, $\text{FPR} = 14.3\%$), while DataSentinel-L reaches $43.5\%$ accuracy and $43.3\%$ macro-F1 ($\text{FNR} = 75.1\%$, $\text{FPR} = 0.8\%$). 
Because tool-poisoning payloads are crafted to blend seamlessly into legitimate tool usage instructions rather than explicitly overriding a canary-repetition task, known-answer detection frequently fails to trigger on stealthy tool injections.

\subsection{Impact of Backbone LLMs}
\label{sec:backbone_impact}
\begin{table}[htbp]
\centering
\footnotesize
\setlength{\tabcolsep}{3.5pt}
\begin{tabular}{lcccccc}
\toprule
Model & Accuracy & Precision & Recall & F1 & FNR & FPR \\
\midrule
Qwen3-8B   & 92.1 & 90.9 & \textbf{91.5} & 91.1 & 7.4  & 9.5  \\
Qwen3-14B  & \textbf{93.8} & \textbf{94.1} & 91.1 & \textbf{92.5} & \textbf{3.5}  & 12.7 \\
Mistral-7B & 53.7 & 63.4 & 63.6 & 55.1 & 56.2 & 15.1 \\
Phi-4      & 88.3 & 92.2 & 80.3 & 84.8 & 3.7  & 31.0 \\
Vicuna-7B  & 23.1 & 56.7 & 44.2 & 21.8 & 98.1 & \textbf{0.8}  \\
\bottomrule
\end{tabular}
\caption{Detection performance (\%) of our method with different backbone models on 503 tools (377 prompt-injected, 126 benign). Precision, recall, and F1 are macro-averaged over the prompt-injected and benign classes. FNR: fraction of injected tools not flagged; FPR: fraction of benign tools wrongly flagged.}
\label{tab:overall-results}
\end{table}

Among the evaluated backbones, \textbf{Qwen3-14B} achieves the strongest overall performance, reaching $93.8\%$ accuracy and $92.5\%$ macro-F1. \textbf{Qwen3-8B} performs comparably at $92.1\%$ accuracy and $91.1\%$ macro-F1, while generating fewer false alarms on benign tools ($\text{FPR} = 9.5\%$ vs.\ $12.7\%$ for Qwen3-14B). 
In contrast, \textbf{Phi-4} ($88.3\%$ accuracy, $84.8\%$ macro-F1) matches Qwen3-14B in low false-negative rate on injected tools ($\text{FNR} = 3.7\%$ vs.\ $3.5\%$) but exhibits over-flagging behavior, misclassifying $39$ of the $126$ benign tools ($\text{FPR} = 31.0\%$).
\textbf{Mistral-7B} ($53.7\%$ accuracy, $55.1\%$ macro-F1) exhibits the opposite bias, defaulting toward benign predictions ($\text{FPR} = 15.1\%$) and missing more than half of all injected tools ($43.8\%$ injected recall; $\text{FNR} = 56.2\%$). 
Finally, \textbf{Vicuna-7B} performs worst ($23.1\%$ accuracy, $21.8\%$ macro-F1), detecting only $7$ of the $377$ injected tools ($1.9\%$ injected recall; $\text{FNR} = 98.1\%$). This degradation stems from architectural and instruction-following limits: its $2{,}048$-token context window prevented evaluation on $75$ tools, and $65$ outputs failed JSON schema parsing; all unparseable or truncated instances were counted as incorrect predictions.

\subsection{Error Analysis}
To better understand the limitations of code-to-description consistency checking, we manually inspected all 31 misclassified and unparseable instances (6.16\% of the 503 test cases) across best performed primary model, Qwen3-14B.

\pbold{False Positives.}
Across the $N_{\text{benign}} = 126$ benign controls, the model produced 15 false positives ($\text{FPR} = 12.7\%$). Manual inspection identified two causes:
\begin{itemize}
    \item \textbf{Under-Specified Implementations (14 cases):} When a benign description promised rich capabilities (e.g., phrasing such as \textit{``detailed tool or analysis''}), the verifier expected corresponding analytical logic in the source code. In benchmark entries where the underlying implementation merely formatted or simulated the output for simplicity, the model flagged the gap as a semantic mismatch.
    \item \textbf{Reasoning Loops (1 case):} On complex or ambiguous inputs, reasoning-capable models entered prolonged iterative comparisons between the code and description without converging on a final JSON object before hitting the token limit. Consistent with our evaluation protocol, these unparseable (\texttt{None}) responses were recorded as incorrect predictions.
\end{itemize}

\pbold{False Negatives.}
Across the $N_{\text{attack}} = 377$ poisoned instances, the model missed 16 attacks 3.5\% overall FNR), comprising 14 false negatives predicted as \texttt{match} and 2 unparseable (\texttt{None}) outputs:
\begin{itemize}
    \item \textbf{Overlooked Injected Directives (14 cases):} In the majority of missed attacks, the model failed to surface the embedded adversarial payload during comparison, leaving no mention of the extraneous instruction in the generated \texttt{implementation\_summary} or \texttt{mismatch\_reason} fields and erroneously classifying the tool as a \texttt{benign}.
    \item \textbf{Payload Hijacking (\texttt{None} Outputs, 2 cases):} In two test cases containing aggressive directive-override payloads, the injected instructions partially hijacked the model's execution flow—causing it to deviate from the required JSON schema and emit an unparseable (\texttt{None}) response rather than a structured \texttt{injected} verdict.
\end{itemize}

\section{Limitations and Future Work}

\subsection{Limitations}
\pbold{Source-Code Access and Scope:} Our defense requires access to tool implementations ($C$ and $H$), limiting its applicability to closed-source or remote MCP servers that expose only metadata. Moreover, our pipeline verifies semantic consistency rather than inherent safety: a malicious tool whose description transparently advertises harmful behavior is classified as a \texttt{match}.

\pbold{Adaptive Attacks and Model Dependence:} Because auditing relies on an LLM, adaptive attackers could embed deceptive comments or obfuscated logic in the source code to mislead the verifier---a threat vector outside our current scope. Detection efficacy also depends heavily on backbone reasoning capacity, ranging from $\sim$94\% accuracy with Qwen3-14B down to 23\% with Vicuna-7B.

\pbold{Evaluation Constraints:} Existing baselines (PromptArmor, DataSentinel, LLM-Based Detection, and Perplexity Filter) evaluate only natural-language descriptions rather than code, creating an asymmetric comparison. Additionally, legitimate instructional metadata can trigger false alarms, and our public-repository dataset exhibits a class imbalance (377 poisoned vs.\ 126 benign tools, where an all-positive baseline achieves $\sim$75\% accuracy) that may not capture the full diversity of production MCP ecosystems.

\subsection{Future Work}
\pbold{Layered Defenses and Fine-Tuning:} Combining lightweight description-level filters with code-to-description auditing in a staged pipeline could balance precision and recall, while fine-tuning compact models on code-alignment tasks would reduce deployment latency.

\pbold{Continuous Auditing and Broader Benchmarks:} Integrating automated re-verification when tool definitions change would mitigate post-deployment tampering (rug-pull attacks). Future evaluations should also expand to multi-language production MCP servers, adaptive code-obfuscation attacks, and larger frontier models.

\section{Conclusion}
In this work, we propose a pre-execution defense that leverages LLMs to verify whether a tool's natural-language description accurately reflects its underlying code implementation. 
We evaluated our approach on a benchmark of 503 benign and poisoned tools against four existing prompt-injection defenses: PromptArmor, DataSentinel, LLM-Based Detection, and Perplexity Filter. 
Experimental results demonstrate that incorporating source-code context significantly strengthens detection against tool poisoning, achieving 93.8\% accuracy and outperforming baselines. 
While these baselines remain effective in the text-level scenarios for which they were originally designed, our findings highlight the importance of cross-checking metadata against executable logic in tool-integrated environments. 
Future work will explore layered defense architectures that ensemble description-level filters with code-consistency auditing, alongside evaluating additional backbone models and expanding benchmark coverage across diverse MCP ecosystems.

\bibliography{main}

\end{document}